\documentclass[article]{revtex4}
\usepackage{graphicx}
\usepackage{amsmath,amssymb}
\usepackage{epsfig}
\usepackage{amsfonts, color, soul}

\newcommand \be  {\begin{equation}}
\newcommand \beq {\begin{equation}}
\newcommand \bea {\begin{eqnarray} \nonumber }
\newcommand \ee  {\end{equation}}
\newcommand \eeq {\end{equation}}
\newcommand \eea {\end{eqnarray}}
\newcommand{\beqa}{\begin{eqnarray}}
\newcommand{\eeqa}{\end{eqnarray}}

\newcommand{\avg}[1]{\langle{#1}\rangle}

\newcommand{\equil}[2]{\underset{#2}{\overset{#1}{\rightleftarrows}}}

\newcommand{\rme}{\text{e}}
\newcommand{\rmi}{\text{i}}

\begin{document}

\begin{center}
{\bf \large The switching dynamics of the bacterial flagellar motor}\\
{\bf Supporting Information}\\ 
Siebe B. van Albada, Sorin T\u{a}nase-Nicola and Pieter Rein ten Wolde
\end{center}

\vspace*{0.5cm}

\tableofcontents

\vspace*{0.5cm}

In this \emph{Supporting information} we provide background
information on our model of the bacterial flagellar motor. We also derive the
analytical solution of our coarse-grained model of the switching
dynamics and explain the hybrid stochastic algorithm used for the
simulations.

\section{The model of the bacterial flagellar motor}

\subsection{Stator-Rotor interaction}
The model for the stator-rotor interaction is discussed in the
sections {\em The stator-rotor interaction} and {\em The rotor
  switching dynamics} of the main text. The model is based on the
model of Oster and Blair and coworkers \cite{Xing:2006dq,Kojima01},
but extended to include the conformational transitions of the rotor protein
complex. Here, we discuss aspects of the model that are not discussed
in the main text. But, for completeness, we also give the main
equations already presented in the main text.

In our model, each stator-rotor interaction is described by 4 energy surfaces,
$U_{s_j}^{r}$, with the subscript $s_j=0,1$ denoting the conformationals
state of stator protein $j$ and the supersript $r=0,1$ denoting the
conformational state of the rotor (clockwise or counterclockwise). We
assume that the stator proteins are fixed by the peptidoglycan layer
and that only the rotor complex moves. The equation-of-motion of the rotor is
then given by
\begin{equation}
\label{eq_SI:Lan}
\gamma_{\rm R} \frac{d\theta_{\rm R}}{dt} = -\sum_{j=1}^{N_{\rm S}}\frac{\partial
  U^r_{s_j}(\theta_j)}{\partial \theta_{\rm R}} +
  F_{\rm L} + \eta_{\rm R} (t).
\end{equation}
Here, $\gamma_{\rm R}$ is the friction coefficient of the rotor;
$U^r_{s_j} (\theta_j)$ is the free-energy surface shown in Fig. 2 of the
main text, where $\theta_j = \theta_{\rm R} -\theta_{{\rm S}_j}$, with
$\theta_{\rm R}$ the rotor rotation angle and $\theta_{{\rm S}_j}$ the
fixed angle of stator protein $j$; $\eta_{\rm R}(t)$ is a
Gaussian white noise term of magnitude $\sqrt{2k_{\rm B}T \gamma_{\rm
    R}}$; $N_{\rm S}$ is the number of stator proteins. The torque
$F_{\rm L}$ denotes the external load. As discussed in
\cite{Elston00,Elston00_2,Xing:2006dq}, for the system studied here,
the torque-speed curves under conservative load and viscous load are
identical. However, as discussed in the main text, the type of load does
markedly affect the CW $\leftrightarrow$ CCW switching dynamics.

The transition (or {\em hopping}) rate for a stator protein to go from
one energy surface to another depends upon the free-energy barrier
separating the two surfaces. We make the
natural phenomenological assumption that the hopping rate depends
exponentially on the free-energy difference, in a manner that obeys
detailed balance. Furthermore, following Blair and Oster and coworkers,
we assume that the access of the periplasmic protons to the
stator-binding sites is triggered by a rotor-stator interaction
\cite{Kojima01,Xing:2006dq}. This yields the following expression for the
hopping rates:
\begin{equation}
k^r_{s_j\to s_j^\prime}(\theta_j)=k_0w(\theta_j)\exp[\Delta U_{ss^\prime}(\theta_j)/2], \,\,\,s,
s^\prime = 0, 1.
\label{eq_SI:k_hop}
\end{equation}
Here, $k_0$ sets the basic time scale, and $\Delta U_{ss^\prime} (\theta_j)
= U_{s^\prime}(\theta_j) - U_{s}(\theta_j)$. The function $w(\theta_j)$
describes the proton hopping windows (see
Fig. 2 of the main text), which reflect the idea that the ion
channel through the stator is gated by the motion of the rotor.

The rotor complex is modeled as an MWC model \cite{Monod65}, which
means that all the rotor proteins switch conformation in concert. This
leads to the following expression for the instantaneous {\em switching} rate:
\begin{equation}
\label{eq_SI:k_s_r}
k^{r\to r^\prime} (\{\theta_j\})=\tilde{k}_0\exp[\Delta U^{rr^\prime}(\{\theta_j\})/2],
\,\,r, r^\prime = 0,1,
\end{equation}
where $\Delta U^{rr^\prime}(\{\theta_j\})=\sum_{j=1}^{N_{\rm S}}
U_{s_j}^{r^\prime} (\theta_j) -U_{s_j}^r(\theta_j)$.  The average,
effective switching rate is given by
\begin{equation}
\label{eq_SI:k_switch}
k_{\rm switch}^{r \to r^\prime} = \int d\theta_{\rm R} P (\theta_{\rm R}) k^{r\to
  r^\prime}_{s_j}(\{\theta_j\}),
\end{equation}
where $P(\theta_{\rm R})$ is the stationary distribution of the
rotor's position. The instantaneous switching rate $k^{r\to
  r^\prime}_{s_j}(\{\theta_j\})$ does not depend upon the
load. Indeed, in our model, the load does not directly affect the
probability that the rotor proteins switch conformation. In this
respect, the mechanism that we propose differs fundamentally from that
often used to explain the force dependence of processes such as
protein unfolding and molecular dissociation \cite{Howardbook}; in
that mechanism one assumes that the reaction coordinate can described
by a single order parameter, and that the force directly couples to
that coordinate, changing the relative stability of the two
(meta)stable states, as well as the location and stability of the
transition state separating them.  In our model, the propensity for
the rotor to switch depends on interactions with the stator
proteins. Consequently, the reaction coordinate for switching depends
not only on the coordinate describing the conformational state of the
rotor protein complex, but also on the coordinates describing the
positions and the conformational of the stator proteins. While the
load may change the free-energy landscape in the direction describing
the conformational state of the rotor, we assume that the load only
couples to the rotation direction of the rotor. The load thus changes
the steady-state distribution of the rotor's position relative to that
of the stator proteins, which in turn affects how often
during their motor cycle the stator proteins favor one conformational
state of the rotor protein complex over the other. In other words,
while increasing the load does not change the instantaneous switching rate
$k^{r\to r^\prime}_{s_j}(\{\theta_j\})$, it does shift $P(\theta_{\rm
  R})$ to positions $\theta_{\rm R}$ where $k^{r\to
  r^\prime}_{s_j}(\{\theta_j\})$ is large.  This is the principal
mechanism that, according to our model, makes the effective switching
rate $k_{\rm switch}^{r\to r^\prime}$ sensitive to
load and speed.\\
\noindent{\bf \sf The load} In the experiments of Korobkova {\em et al}. the motion of the flagellum
is visualized via a latex bead connected to the flagellar filament
\cite{Korobkova04,Korobkova06}. The bead exerts a force on the rotor
protein, which, effectively, tilts the energy surfaces shown in
Fig. 2 of the main text. When a) the connection between the load
and the motor is soft, b) the dynamics of the motor is much faster
than that of the load, and c) chemical transitions lead on average to
a fixed translation distance of the rotor, as in the current model,
then the torque-speed curves under conservative load and viscous load
are identical \cite{Elston00,Elston00_2,Xing:2006dq}. However, as
discussed in the main text, the type of load does markedly affect the CW
$\leftrightarrow$ CCW switching dynamics.

\begin{figure}[t]
\includegraphics[width=7cm]{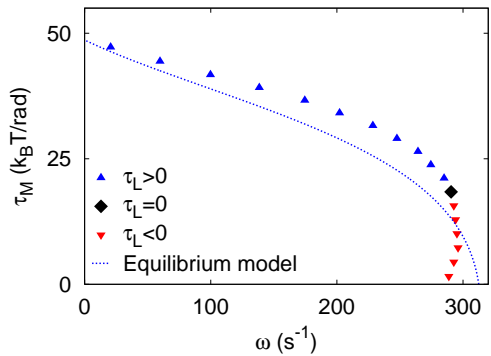}
\caption{\label{fig_SI:torque-speed} The torque-speed relation (symbols)as
  predicted by the model used here, which is based on the model of
  Oster {\em et al.} \cite{Xing:2006dq}.  The
blue symbols correspond to the regime in which the load pulls the
motor in the backward direction; the red symbols correspond to the
regime in which the load pulls the motor in the {\em forward}
direction, which is the scenario right after a switching event.  The
line shows that, to a
good approximation, the speed $\omega$ as a function of the motor
torque $\tau_{\rm M}$, is given by the speed as a function of the
conservative load $\tau_{\rm L}$, according to
$\omega = (k_{\rm hop}^+-k_{\rm hop}^-) \pi/26$, where the hopping
rates in the forward (+) and backward (-) are given by 
 $k_{\rm
  hop}^\pm =\int_0^{2\pi/26}d\theta P_{\rm S} (\theta) k^\pm(\theta)$, and the stationary
distribution $P_{\rm S}(\theta)$ is approximated by the equilibrium distribution
 $P_{\rm eq}(\theta) \propto
\exp(-\beta\left[U_{s}(\theta)+\tau_{\rm L} \theta\right])$ (See
Ref. \cite{ThesisVanAlbada}).
The
parameters used in the simulations are shown in Table \ref{tab_SI:1}.}
\end{figure}

\noindent{\bf \sf The stator proteins} Resurrection experiments
suggest that {\em in vivo} the number of stator proteins is around
$8-12$ \cite{Ryu00,Reid06}. At high load, the stator proteins act
cooperatively, and the motor speed increases with the number of stator
proteins \cite{Ryu00}; the model of Oster and coworkers describes this
observation \cite{Xing:2006dq}. Recent experiments by Yuan Berg show
that near zero external load, the speed is independent of the number
of stator proteins \cite{Yuan08}. The model of Oster and coworkers can
reproduce this behavior if the stator proteins are connected to the
rigid framework of the cell wall via very soft springs. However, to
generate a speed that is independent of the number of stator proteins,
the springs have to be made so soft that they stretch a distance of
order 10 nm, which, as Yuan and Berg point out, seems unlikely
\cite{Yuan08}. We therefore focus here on a motor that has only one
stator protein that is rigidly connected to the cell membrane. This
motor has a lower maximum torque than a ``wild-type''motor with 8-12
stator proteins, but this is not critical, since we take a rather
small bead (see table \ref{tab_SI:1}); in essence, to a good
approximation, all the torques in our model could be scaled by the
number of stator proteins. More importantly, our model correctly
predicts the maximum speed of about 300 Hz, as recently observed by
Yuan and Berg \cite{Yuan08}, and its torque-speed relation exhibits
the distinc knee at a speed of about 250 Hz (see
Fig. \ref{fig_SI:torque-speed}). This model thus captures the effect of
the dynamical interplay between the torque and the speed, and on the
other hand the switching dynamics. The maximum speed is particularly
important, since that, together with the total change in the winding
angle of the flagellar filament upon a motor reversal, directly
affects the characteristic switching time. In future work, we will
investigate the effect of the number of stator proteins on the
switching dynamics.

The parameters of the rotor-stator model, which were mostly taken from
\cite{Xing:2006dq}, are summarized in table \ref{tab_SI:1}.

\begin{table}[b]
\begin{center}
\begin{tabular}{lrr} \hline\hline\\[-0.4cm]
 Parameter   & Value & Description \\ \hline
$d$        & $2\pi~{\rm rad}/26$     & Potential periodicity\\
pmf        & 152 mV            & Proton-motive force              \\
$\Delta G$ & $11.8~ k_{\rm B}T$ & $\Delta G = 2e \times {\rm pmf}$ \\
$k_0$        & $3.5~ 10^4~{\rm s}^{-1}$& Hopping prefactor                 \\
$d_1$      & $0.05~d$          & Position potential maximum \\
$d_2$      & $0.1~d$           & Position start power stroke\\
$d_3$      & $0.9~d$           & Center of hopping window   \\
$d_4$      & $0.2~d$           & Width of hopping window   \\
$h_1$      & $25~k_{\rm B}T$         & Height potential maximum\\
$h_2$      & $10~k_{\rm B}T$         & Height start power stroke\\
$F_{\rm M} = -h_2 / (d - d_2)$ & $46~ k_{\rm B}T~{\rm rad}^{-1}$& Force motor during power stroke\\
$\gamma_{\rm M}$ & $1.7~10^{-3}~{\rm s}~{\rm rad}^{-2}$  & Friction coefficient motor\\
$\tilde{k}_0$  & $0.3~{\rm s}^{-1}$  & Switching prefactor\\
$\gamma_{\rm L}$ & $0.51 k_{\rm B}T~{\rm s}~{\rm rad}^{-2}$  & Friction coefficient load\\
\end{tabular}
\end{center}
\vspace{0.3cm}
\caption{\label{tab_SI:1} Parameters for the rotor-stator model as used
  in the simulations (see also \cite{Xing:2006dq}).}
\end{table}

\subsection{Elasticity of the flagellar filament}
 We assume
that the free energy of a flagellar filament in a given polymorphic state $m$ is
quadratic in the curvature $\kappa$ and torsion $\tau$:
\begin{equation}
U^{\rm F}_m(\tau,\kappa)/L = \frac{1}{2}EI (\kappa-\kappa_m)^2 +
\frac{1}{2}\mu J(\tau-\tau_m)^2,
\end{equation}
where $L$ is the contour length, $E$ and $\mu$ are the Young's and
shear moduli, $I$ and $J$ are cross-sectional moments, and $\kappa_m$
and $\tau_m$ are, respectively, the spontaneous curvature and torsion
of the filament in state $m$. The curvature and torsion are functions
of the height $z$ and the winding angle $\theta$:
\begin{eqnarray}
\kappa (\theta,z)&=&\frac{\theta\sqrt{L^2-z^2}}{L^2}, \\
\tau(\theta,z)&=&\frac{\theta z}{L^2}.
\end{eqnarray}

 We assume that at
each instant the length of the filament has relaxed to its steady state
value $z_{\rm eq}(\theta)$, obtained as a solution of the equation
 $\frac{\partial U(\tau(\theta,z),\kappa(\theta,z))}{\partial z} =0$. This allows us to elimate $z$ and express the ``torsional'' energy as a function of the winding angle $\theta$:
\begin{equation}
U^{\rm T}_m(\theta) = U^{\rm F}_m(\tau(\theta,z_{\rm eq}(\theta)),\kappa(\theta,z_{\rm eq}(\theta))),
\end{equation}
The function $U^{\rm F}_m(\theta)$ is, in general, a complicated fonction of $\theta$; nevertheless in the limit of equal 
bending and twisting stifnesses ($E I=\mu J$)~\cite{Darnton:2007hc}  the torsion potential corresponds to a linear elastic potential 
\begin{equation}
U^{\rm F}_m(\theta) = \frac{1}{2} k_\theta (\theta - \theta_m)^2,
\end{equation}
where $k_\theta=\frac{E I}{L}$ and
$\theta_m=\frac{\sqrt{\kappa_m^2+\tau_m^2}}{L}$. The eperimental data
of Darnton and Berg \cite{Darnton:2007hc} confirm that the
approximation $E I\simeq\mu J$ is valid and therefore, locally, the
potential energy guiding the dynamics of the twisting angle $\theta$
has a simple linear elasticity form with elastic constant $k_m \simeq
100$ pN nm/${\rm rad}^2$ (obtained from $EI=\mu J=3.5$ pN ${\rm
  \mu} m^2$ and $L=7.6, 19.6$ ${\rm \mu}$m as in
\cite{Darnton:2007hc}). As described in the main text, we assume that
the potentiall wells are equally spaced, are of the same depth and
have the same curvature. Clearly, these assumptions could be relaxed
by allowing, {\em e.g.}, the normal state to be more stable and to have a
higher stiffness. 

Motivated by the observations of Darnton and Berg \cite{Darnton:2007hc}, we assume that the transition from one
polymorphic state to another is an activated process, with a rate
constant given by
\begin{equation}
k_{m\to m^\prime}(\theta) = \breve{k}_0 \exp[(U^{\rm
  F}_m(\theta)-U^{\rm F}_{m^\prime}(\theta))/2].
\label{eq_SI:k_p}
\end{equation}

The equation-of-motion for the bead is given by
 \begin{equation}
\gamma_{\rm L} \frac{d\theta_{\rm L}}{dt}= -k_\theta
(\theta_{\rm L} - \theta_{\rm R} -\theta_{m}) + \eta_{\rm L} (t).
\label{eq_SI:viscous_load}
\end{equation}
Here, $\gamma_{\rm L}$ is the friction coefficient of the bead, and
$\eta_{\rm L}$ is a Gaussian white noise term of magnitude $\sqrt{2
  k_{\rm B}T \gamma_{\rm L}}$. 

The parameters of the model are given in table \ref{tab_SI:2}.

\begin{table}[t]
\begin{center}
\begin{tabular}{lrr} \hline\hline\\[-0.4cm]
 Parameter   & Value & Description \\ \hline
$\theta_m-\theta_{m-1}$        & $ 150 d$     &
Spacing of wells\\
$k_\theta$        & $1 k_{\rm B}T /{\rm rad}^2$            & Stiffness\\
$N$ & 10 & Number of wells.\\
$\breve{k}_0$&$10^{-6}{\rm s}^{-1}$&Jumping prefactor
\end{tabular}
\end{center}
\vspace{0.3cm}
\caption{\label{tab_SI:2} Parameters describing the flagellar filament.}
\end{table}

\section{Coarse grained model of the switching dynamics}

We  model the switching dynamics as a memoryless 
two-state system with switching-time distributions
$\psi_{+}(t)$ switching from CW to CCW and $\psi_{-}(t)$ for switching from CCW to CW:
\beq
\mathrm{CW}\equil{\psi_{+}(t)}{\psi_{-}(t)}\mathrm{CCW}
\eeq
Lack of memory means in this context that the probability to switch  from one state depends only on the time
since the transition to that state -- the system 
forgets everything before the last transition.

The switching-time distribution is related to the switching rate or
switching propensity (the switching probability per unit amount of
time) $k_{\alpha}(t)$ as
 \beq
\psi_{\alpha}(t)=k_{\alpha}(t)\rme^{-\int_0^t k_{\alpha}(t^\prime) dt^\prime}.
\eeq

One important characteristic of the stochastic trajectory of the
system is the correlation function $C(t)$ of the characteristic
function $\chi(t)$:
 \beq
C(t)=\avg{\chi(t)\chi(0)}-\avg{\chi}^2. \eeq 
We take $\chi(t)=1$ if the system is in the CW
state and $\chi(t)=0$ otherwise. From the ensemble of all
possible trajectories only the ones that are in the CW state both at
time zero and at time $t$ contribute to the correlation function at
time $t$. Therefore, one can write the correlation function as \beq
C(t)=\left[P({\rm CW},t;{\rm CW},0)-P({\rm CW},\infty;{\rm
    CW},0)\right]P({\rm CW},\infty;{\rm CW},0), \eeq where $P({\rm
  CW},t;{\rm CW},0)$ is the probability that a trajectory is in the CW
state at time $t$ given that it starts in that state at time zero.
Using a well established result in the theory of (alternating)
two-state, memoryless renewal processes (see \cite{Cox:1961eu},
Chapter 7) one can express this quantity in the Laplace domain as:
\beq \tilde P(z)=\frac{1}{z}\left(1-\frac{G(z)}{z t_{\rm
      CW}}\right). \eeq Here, $\tilde{P}(z)$ is the Laplace transform
of $P({\rm CW},t;{\rm CW},0)$, \beq \tilde P(z)=\int_0^\infty P({\rm
  CW},t;{\rm CW},0)\rme^{- z t} dt, \eeq $G(z)$ is a function that
depends on the Laplace transformed switching-time distributions, \beq
G(z)=\frac{(1-\tilde \psi_{+}(z))(1-\tilde \psi_{-}(z))}{(1-\tilde
  \psi_{-}(z)\tilde \psi_{+}(z))}, \eeq and $t_{\rm CW}$ is the
average residence time in the CW state.  The probability to be in the
CW state is given by the average residence times as \beq P({\rm
  CW},\infty;{\rm CW},0)=\frac{t_{\rm CW}}{t_{\rm CW}+t_{\rm CCW}}.
\eeq Also, using the properties of the Laplace transform one has \beq
\tilde C(z) = \frac{t_{\rm CW}}{t_{\rm CW}+t_{\rm CCW}}\left[ \tilde
  P(z)- \frac{t_{\rm CW}}{z(t_{\rm CW}+t_{\rm CCW})}\right]. \eeq Once
we have the correlation function, we can compute the power spectrum
using the formula \beq S(\omega)=2 \int_0^\infty C(t)\cos{\omega
  t}=\tilde C(\rmi \omega)+\tilde C(-\rmi \omega), \eeq such that \beq
S(\omega)=\frac{1}{\omega^2(t_{\rm CW}+t_{\rm CCW})}\left[G(\rmi
  \omega)+G(-\rmi \omega)\right]. \eeq
\begin{figure}[t]
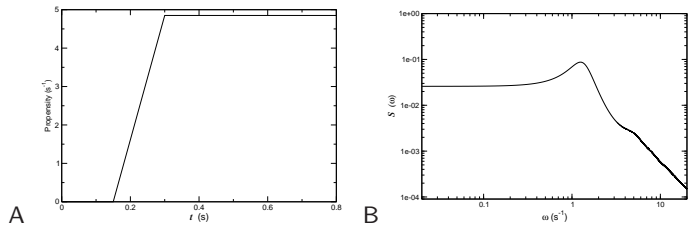

{\bf \sf A}\,\, \includegraphics[width=4cm]{Fig_k}\,\,\,
{\bf \sf B} \includegraphics[width=4cm]{Fig_S}
\caption{\label{fig_SI:model} A) A piecewise linear model of the
  switching-propensity function $k(t)$. B) Computed power spectrum $S(\omega)$.  }
\end{figure}

In general, an analytical formula for the power spectrum $S( \omega)$
cannot be obtained for any arbitrary switching-propensity function
$k_\alpha(t)$.  Nevertheless, one can obtain an analytical formula for
the power spectrum if the switching-propenstiy function is piecewise
linear, as in Fig. \ref{fig_SI:model}A. Fig. \ref{fig_SI:model}B shows the
power spectrum for a symmetric system, with switching-propensity
functions in the forward and backward directions as shown in
Fig. \ref{fig_SI:model}A. It is seen that this simple, non-Markovian two-state
model can capture the main features of the power spectrum as measured
by Korobkova {\em et al.}\cite{Korobkova06}.

\section{Hybrid stochastic algorithm}
The equations-of-motion for the rotor and the flagellum,
Eqs. 1 and 7 of the main text, respectively, and Eqs. \ref{eq_SI:Lan} and
\ref{eq_SI:viscous_load} above, are propagated via a Heun scheme
\cite{Greiner88}. 

The algorithm to determine when the next hopping, switching, or
polymorphic transition will occur is essentially a kinetic Monte Carlo
algorithm \cite{Bortz75}. It is based on the observation that the {\em
  survival} probability $S(t)$, i.e. the probability that no hopping,
switching or polymorphic transition has happened after a time $t$
after the last event, is given by
\begin{equation}
S(t) = \exp\left(-a(t)\right),
\end{equation}
where $a(t)$ is the cumulative total propensity function:
\begin{equation}
a(t) = \int_0^t dt^\prime k_{\rm
  T}(t^\prime),
\end{equation}
 with $k_{\rm T}(t)$ being the total propensity function as given
by
\begin{equation}
k_{\rm T}(t) = \sum_{j=1}^{N_{\rm S}}k^r_{s_j\to s_j^\prime}(\theta_j(t)) +k^{r\to
  r^\prime} (\{\theta_j(t)\}) + k_{m\to m^\prime}(\theta_{\rm
  L}(t)-\theta_{\rm R}(t)-\theta_m)).
\end{equation}

In practice, right after a hopping, switching or polymorphic
transition, a random number, $\xi$, between zero and one is drawn. The
equations-of-motion of the rotor and the flagellum are then integrated
together with the equation that describes the temporal evolution of
$a(t)$:
\begin{equation}
\label{eq_SI:dadt}
\frac{d a(t)}{dt} = k_{\rm T}(t).
\end{equation}
Integrating Eq. \ref{eq_SI:dadt} since the last event leads to an
estimate for $a(t) = \int_0^t dt^\prime k_{\rm T}(t^\prime)$.
The next event
then occurs after a time $t$ since the last event when
\begin{equation}
a(t) > \log(1/\xi).
\end{equation}
 The event type $\alpha$, where $\alpha$ is
either a hopping, switching, or polymorphic transition, is subsequently chosen
with a probability $p_{\alpha}$ as given by
\begin{equation}
p_\alpha (t) = k_\alpha(t) / k_{\rm T}(t).
\end{equation}

\bibliography{/Users/tenwolde/articles/references/sysbio.bib}

\end{document}